\documentclass[aps,prl,twocolumn,superscriptaddress,letterpaper,preprintnumbers,nofootinbib,noshowpacs]{revtex4}
\pdfoutput=1
\usepackage{graphicx,hyperref}
\newcommand{\vev}[1]{\langle {#1} \rangle}
\newcommand{\lsim}{\lesssim}
\newcommand{\gsim}{\gtrsim}

\newcommand{\eq}[1]{Eq.~(\ref{#1})}
\newcommand{\ord}[1]{\mathcal{O}{(#1)}}
\newcommand{\beq}{\begin{equation}}
\newcommand{\eeq}{\end{equation}}
\newcommand{\eps}{\varepsilon}

\begin{document}

\pagestyle{plain}

\title{Baryon Number Violation via Majorana Neutrinos\\ 
in the Early Universe, at the LHC, and Deep Underground}

\author{Hooman Davoudiasl}
\affiliation{Department of Physics, Brookhaven National Laboratory, Upton, NY 11973, USA}
\author{Yue Zhang}
\affiliation{Walter Burke Institute for Theoretical Physics,
California Institute of Technology, Pasadena, CA 91125, USA 
\vspace{0.1in}\\
{\tt hooman@bnl.gov\ \ \ \ \ yuezhang@theory.caltech.edu}
}


\begin{abstract}

We propose and investigate a novel, minimal, and experimentally testable framework for baryogenesis, dubbed {\it dexiogenesis}, using baryon number violating effective interactions of right-handed Majorana neutrinos responsible for the seesaw mechanism. The distinct LHC signature of our framework is same-sign top quark final states, possibly originating from displaced vertices.  The region of parameters relevant for LHC phenomenology can also yield concomitant signals in nucleon decay experiments.  We provide a simple ultraviolet origin for our effective operators, by adding a color-triplet scalar, which could ultimately arise from a grand unified theory.

\end{abstract}

\preprint{CALT-TH-2015-023}

\maketitle

Gauge singlet right-handed neutrinos (RHNs) provide perhaps the simplest explanation of non-zero neutrino masses, as demanded by a large and well-established body of neutrino oscillation data, thereby allowing the Standard Model (SM) to be a renormalizable theory of Nature.  
If neutrinos are Dirac particles, the associated RHNs only couple to the SM via Yukawa interactions of negligible strength, $y_N \lsim 10^{-12}$, and are not expected to be directly detectable in the foreseeable future. On the other hand, if the observed neutrinos are Majorana states, they most naturally get their mass from a seesaw mechanism~\cite{seesaw}. In that case, RHNs may have $\ord{1}$ couplings to SM neutrinos, but they would then have to be exceedingly heavy, $\gsim 10^{14}$\,GeV, far beyond the reach of terrestrial experiments. However, it may very well be that RHNs are much lighter, and they can be directly probed in lepton number violating processes in collider experiments~\cite{Keung:1983uu, Datta:1993nm, Atre:2009rg, Nemevsek:2011hz, Deppisch:2015qwa} and perhaps other searches, such as those for proton decay~\cite{Davoudiasl:2014gfa}.

In this letter, we assume that the RHNs associated with the seesaw mechanism are near the weak scale $\lsim 1$\,TeV. Higher dimensional operators involving the RHNs are generically present. In particular, we will further assume that the RHNs have baryon number violating interactions, mediated by dimension 6 operators involving right-handed quarks and suppressed by a scale $\sim1-10$~TeV. We will show that these assumptions allow for direct generation of a baryon number asymmetry through RHN decays in the early Universe, which we dub {\it dexiogenesis} (dexios: Greek for the right hand).  This is in contrast to canonical leptogenesis~\cite{Fukugita:1986hr} where the lepton asymmetry needs to be further processed into baryon number through electroweak sphalerons~\cite{nosphaleron}. Our direct baryogenesis mechanism is most constrained by nucleon decay bounds.  However, we show that for viable parameters one could have distinct collider signatures.  This scenario, with dim-6 operators, is an effective field theory and can be embedded in a simple renormalizable model, and possibly a grand unified theory (GUT), where additional signals are expected to arise at the Large Hadron Collider (LHC) and future colliders.  For a partial list of other works whose subjects have some overlap with that of this letter, see, for example, 
Refs.~\cite{Cheung:2013hza,Sierra:2013kba,Baldes:2014rda,Monteux:2014hua}.

Our point of departure is the SM augmented by two Majorana RHNs $N_a$, $a=1,2$, 
the minimum required for a realistic seesaw
mechanism based on current neutrino data.  We add the following terms to the Lagrangian
\beq
{\cal L}_N = M_a \bar N_a^{c} N_a + y_N^{a i} H\bar N_a L_i + \text{\small h.c.}\,,
\label{LN}
\eeq
where $M_a$ is the mass of $N_a$, $i=1,2,3$ enumerates SM generations, and $y^{a i}$ is a Yukawa matrix; 
$H$ and $L_i$ are the Higgs and lepton doublets of the SM, respectively.

Light neutrino masses $m_\nu \lsim 0.1$~eV, implied by the
oscillation data, can be generated from the renormalizable interactions in \eq{LN}, 
via the seesaw mechanism: $(m_\nu)_{ij} \sim y_N^{ai} y_N^{aj} \vev{H}^2/M_a$.  Nonetheless,
the SM, henceforth defined to include the Lagrangian in \eq{LN},
is widely expected to be an effective theory that is further enriched with new interactions at higher scales.
This expectation is strongly motivated by the need for a dark matter candidate and also a baryogenesis mechanism to generate
the observed baryon asymmetry of the Universe.

Baryogenesis requires a source of baryon number
violation~\cite{Sakharov:1967dj}. Hence, it may be necessary to extend the SM by 
effective operators that violate baryon number~\cite{Weinberg:1979sa}.  Such operators
are suppressed by scales associated with new physics, often considered to be very high, $\gsim 10^{15}$~GeV,
as implied by nucleon decay constraints.  However, it is compelling to look for scenarios where new physics
arises at lower scales.  For one thing, it is reasonable to assume that the physics underlying the Higgs potential is not
very far from the weak scale.  Such physics would then have the added benefit of being potentially testable.

Motivated by the above considerations, we assume the following baryon-number violating
operators involving the RHNs, in addition to those made up of only the observed SM fields,
\beq
{\cal L}_{\rm BV} = \frac{\lambda_a^{ijk}}{\Lambda^2} [N_a u_i d_j d_k]_R + 
\frac{\kappa_a^{ilm}}{\Lambda^2}[N_a d_i]_R [Q_l Q_m]_L + {\rm h.c.}\,,
\label{LBV}
\eeq
where $i,j,k$ are family numbers of right-handed quark ($u,d$) mass eigenstates and $l,m$ enumerate 
left-handed quark ($Q$) generations.  Here,  $\lambda_a^{ijk}$ and $\kappa_a^{ilm}$ 
are generally complex constants determined by the ultraviolet (UV) theory. 
These operators could arise from grand unified theories, as shown in a concrete example in the end.
They are the lowest dimensional operators that allow RHNs 
to couple to baryon number directly~\cite{otherframework}.

In order to have successful dexiogenesis, 
the coefficients of the relevant dim-6 operators cannot be too small.  
To avoid excessive low energy baryon number 
violation for $\Lambda\gsim 1$~TeV, those operators 
would mainly involve right-handed third generation quarks, 
which help avoid severe constraints from nucleon decay data, 
discussed in more detail below.  To see this, note that quark mass diagonalization can 
induce operators that involve light quarks, in the presence of left-handed fields.   
For this reason, we will not further consider tree-level 
dim-6 operators $NdQQ$ in \eq{LBV}.  These operators    
can still be generated from $Ntbb$ via radiative corrections, 
but would not lead to severe constraints.  

In light of the above discussion, throughout this work, we will focus on operators involving right-handed 
third generation top and bottom quarks (originating from the first term in 
Eq.~(\ref{LBV}), with explicit spinor contractions)
\beq
{\cal L}_{\rm BV}^{\rm 3 R}=
\lambda_a \frac{[\bar N_a^c P_R b][\bar t^c P_R b]}{\Lambda^2}\,,
\label{Ntbb}
\eeq
where $P_R\equiv (1+\gamma_5)/2$ is the right-handed projector.
The dominance of the third generation could be expected from a connection to UV flavor dynamics.  
Operators with other combinations of chiralities and flavors can in principle be present, 
but they must be more suppressed~\cite{exception}.

\begin{figure}[t]
\includegraphics[width=0.42\textwidth]{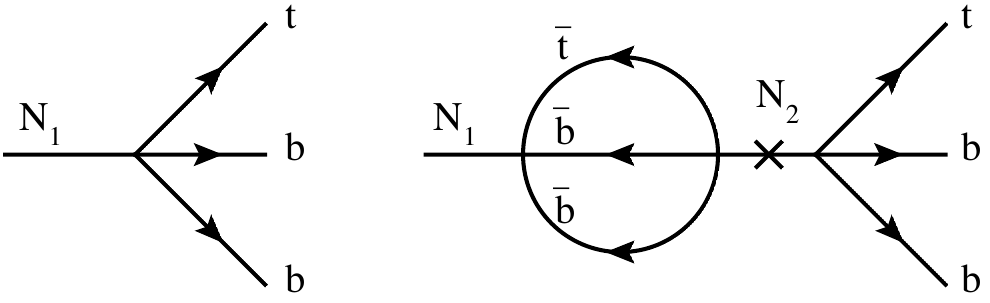}
\caption{Tree and two-loop diagrams for dexiogenesis.}
\label{fig:Genesis}
\end{figure}

Remarkably, the addition of the operators in \eq{Ntbb} provides all the necessary ingredients
encoded in Sakharov's conditions~\cite{Sakharov:1967dj} for baryogenesis: {\it (i)} these interactions
are manifestly baryon number violating, {\it (ii)} their complex coefficients provide a source of CP
violation, and {\it(iii)} if the Universe has a low reheat temperature $T_{\rm RH}\ll M_a$, then
the $N_a$, assumed to be non-thermally produced throughout this work, 
will decay out of equilibrium. This mechanism, dexiogenesis, allows $T_{\rm RH}\ll 100$~GeV,
since the baryon asymmetry is directly generated and hence electroweak sphalerons do not need to be active.  

Let $N_1$ be the lighter of the two RHNs in our setup. Then, the interference of the tree and the 2-loop diagrams in 
Fig.~\ref{fig:Genesis} will lead to a baryon asymmetry
$\eps \equiv {\Gamma(N_1\to tbb) - \Gamma(N_1\to \bar t \bar b \bar b)}/({2 \Gamma_{N_1}})$, 
where the width of $N_1$ is given by 
\beq
\Gamma_{N_1} = \frac{|\lambda_1|^2 M_1^5}{1024 \pi^3 \Lambda^4} F \left( {m_t^2}/{M_1^2} \right)\,,
\label{GamN1}
\eeq
with $F(x) = 1 - 8 x - 12 x^2 \log x + 8 x^3 - x^4$. 

In the presence of the higher dimensional operator \eq{Ntbb} with a TeV scale cutoff, 
$N_1$ decays induced by neutrino Yukawa interactions [\eq{LN}] are subdominant, 
for values of $M_1$ near the weak scale. Given a realistic seesaw mechanism for the SM active neutrino masses, in general we have $y_N^a \lsim 10^{-6} \sqrt{M_1/(\rm 200\,GeV)}$ in the absence of fine tuning~\cite{comment}. The induced $N_1\to W\ell$ decay rate is then estimated to be $\Gamma_{N_1\to W\ell} \lsim 10^{-12}\,{\rm GeV}\, (y_N^a/10^{-6})^2\, [M_1/(\rm 200\,GeV)]$.  We find that, for $M_1$ of a few hundred GeV and $\Lambda/\sqrt{\lambda_1}\lesssim 25\,$TeV, that rate 
is smaller than the baryonic decay rate.

The baryon asymmetry can be conveniently obtained using the unitarity cut method~\cite{Schwartz:2013pla}
\beq
\eps = \frac{\text{Im}(\lambda_1^2 \lambda_2^{*2})}{3072 \pi^3 |\lambda_1|^2} \left(\frac{M_1}{\Lambda}\right)^4
\frac{M_1M_2}{ (M_2^2 - M_1^2)}\,.
\label{eps}
\eeq

The relation between the above asymmetry and the baryon number to entropy ratio $\eta\equiv n_B/s \sim 10^{-10}$~\cite{Agashe:2014kda} 
depends on the non-thermal production mechanism for $N_1$, but it can plausibly be $\eta\sim \eps/100$.  For example, let us assume that 
a heavy modulus, such as an inflaton, decays equally into radiation and $N_1$, which promptly decays.  We will 
take the reheat temperature to be $T_{\text{RH}}\sim 1$~GeV.  Then, one can estimate 
$\eta\sim \eps/g^*$ where $g^*\sim 100$ is the number of relativistic degrees of
freedom at $T_{\text{RH}}$.  Alternatively, if the modulus decays exclusively into $N_1$, and it is the 
decay of the $N_1$ population that reheats the Universe, we end up with $\eta \sim \eps\, T_{\text{RH}}/M_1$, which 
for $M_1 \sim 100$~GeV, again yields $\eta\sim \eps/100$.  Hence, for $M_1\sim M_2$ and $\lambda_a \sim 1$,
we typically require $M_1/\Lambda \gsim 0.1$.  Consequently, for $M_1\lsim 1$~TeV, relevant
for collider phenomenology, the cutoff scale must be sufficiently low, $\Lambda\lsim 10$~TeV.  Let us then
examine the experimental constraints on $\Lambda$.

\begin{figure}[b]
\includegraphics[width=0.3\textwidth]{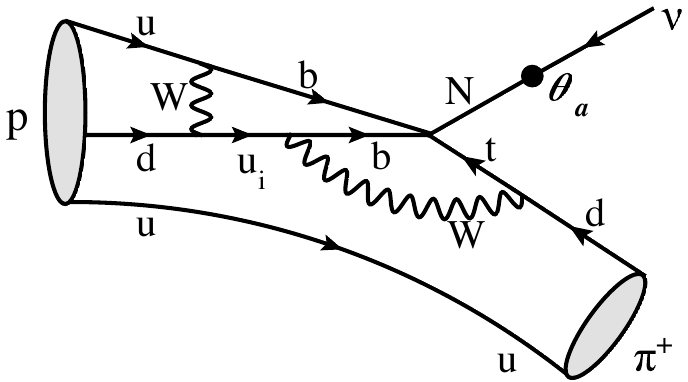}
\caption{One of the leading diagrams that yield proton decay.}
\label{fig:ndecay}
\end{figure}

It turns out that nucleon decay limits provide the most stringent lower bound on $\Lambda$ in the above model 
where RHNs violate both lepton and baryon numbers. While the operators in \eq{Ntbb} do not contain light quarks, quantum loop corrections can induce nucleon decay via these baryon number violating interactions.  Fig.~\ref{fig:ndecay} provides a sample two-loop diagram for proton decay, with a rate given by 
\begin{eqnarray}
\Gamma(p\to \pi^+ \nu) \!=\! \frac{(1+g_A)^2 \alpha^2 m_p}{32\pi f_\pi^2}\! \left| \xi \right|^2\,,
\label{pdecay}
\end{eqnarray}
where $g_A=1.27$ is the nucleon axial charge, $f_\pi=131\,$MeV is pion decay constant, 
lattice calculations~\cite{Aoki:2006ib} yield the form factor $\alpha\approx-0.01125\,$GeV$^3$, and 
\beq
\xi\approx\frac{\Lambda_{qcd} G_F^2 m_t m_b^2 V_{td}^2 V_{ub}^* V_{tb}^*}{(16 \pi^2)^2 \Lambda^2} \lambda_a \theta_{a}
\label{xi}
\eeq
is the Wilson coefficient from estimating the two-loop diagram in Fig.~\ref{fig:ndecay}. 
The angle $\theta_{a}$ is the mixing between $N_a$ and the SM active neutrinos. 
The hadronic mass scale $\Lambda_{qcd}\approx 200\,$MeV must be introduced under a symmetry argument.
The operator we started with is $[\bar N_a^c P_R b][\bar t^c P_R b]$, and after the $W$-loop dressing as in 
Fig.~\ref{fig:ndecay}, the operator for proton decay turns out to be $[\bar N_a^c P_R d][\bar u^c P_L d]$ 
(which is the radiatively generated $NdQQ$ operator mentioned earlier). 
The fact that one of the down quark is still right-handed implies an 
external (constituent) quark mass insertion $\sim \Lambda_{qcd}$.

The resulting proton decay life time is
\beq
\!\!\tau(p\to \pi^+ \nu) \approx 2.5\times 10^{32}\,{\rm yr} \left( \frac{\Lambda/\sqrt{\lambda_a}}{1.5\,{\rm TeV}} \right)^4 \!\left( \frac{\theta_a}{10^{-6}} \right)^{-2} \,.
\label{tau}
\eeq
The current experimental lower limit on the $p\to \pi^+ \nu$ decay 
channel is $1.6\times 10^{31}\,{\rm yr}$ \cite{Agashe:2014kda}.  Hence, the above lifetime  
(\ref{tau}) is not far from the current limit and, in the region of parameters 
considered in our work, can be within the reach of future nucleon decay experiments \cite{NDEXP, Babu:2013jba}.  

Here, we also address a potential bound from requiring that the asymmetry in \eq{eps} 
is not washed out after baryogenesis.  The low reheat temperature 
assumption mentioned before can ensure that processes mediated by the operators  
in \eq{Ntbb}, such as $b b \to N \bar t$, are effectively turned off.  However, loop processes similar 
to those depicted in Fig.~\ref{fig:ndecay} can lead to baryon number violation mediated by lighter states.  Let us 
assume, for illustrative purposes, that the reheat temperature is $T_{\rm RH} \sim 1$~GeV, well above 
the temperature at the onset of Big Bang Nucleosynthesis. We then have to make sure that the analog 
of neutron-anti-neutron oscillation mediated by 
$(c s s) ({\bar c \bar s \bar s})/\Lambda'^5$ 
is not active.   A straightforward comparison with \eq{xi} yields 
\beq
\Lambda'^5 \approx \left( \frac{\theta_a}{\xi} \frac{\Lambda_{qcd}}{m_s} \frac{V_{td}^2 V_{ub}^*}{V_{ts}^2 V_{cb}^*} 
\right)^{2}  M_1 \,,
\label{lam'}
\eeq
where $m_s \approx  100$~MeV is the strange quark mass.  For $M_1 = 200$~GeV, we find 
$\Lambda' \sim 2\times 10^9$~GeV.  The rate of the $c s s \leftrightarrow {\bar c \bar s \bar s}$ process is estimated to be 
$\Gamma_{\Delta B=2}\sim T_{\rm RH}^{11}/\Lambda'^{10} \sim 10^{-93}$~GeV, for $T_{\rm RH}\sim 1$~GeV, 
which is completely negligible compared to the Hubble rate at this temperature, $H\sim T_{\rm RH}^2/M_{\rm planck} \sim 10^{-19}$~GeV.  
We also note that bounds from neutron-anti-neutron oscillation would not constrain our model, since that process 
involves up and down quarks, for which the corresponding suppression scale is larger than the above $\Lambda'$ scale, and much higher than the current limit~\cite{Arnold:2012sd}. 

\begin{figure*}[t]
\includegraphics[width=0.30\textwidth]{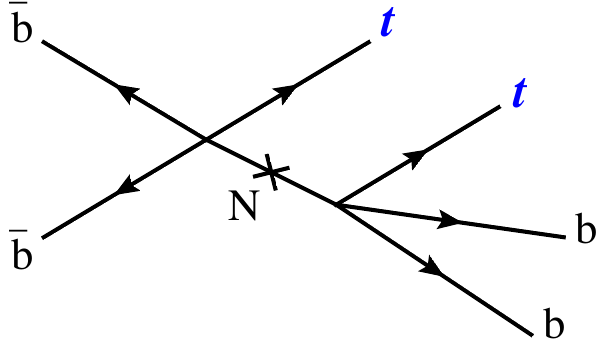}\hspace{0.5cm}
\includegraphics[width=0.33\textwidth]{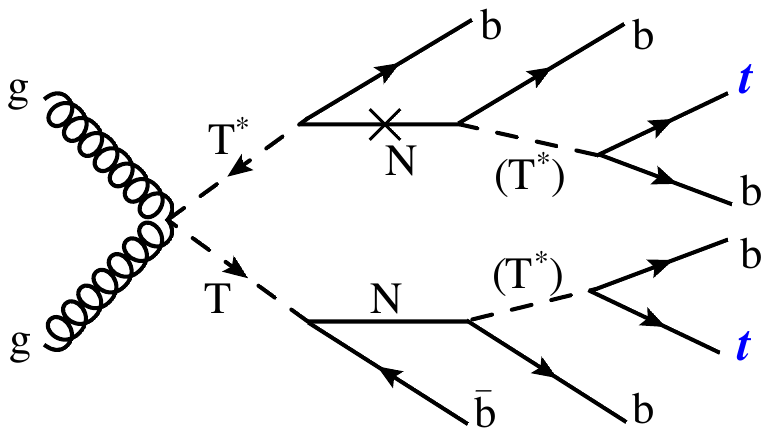}
\caption{Feynman diagrams for same-sign top quark events that can happen at hadron colliders, using the Majorana nature of RHNs and the 
baryonic interactions of Eq.~(\ref{Ntbb}). {\bf Left}: $pp\to t N$ production via the contact operator Eq.~(\ref{Ntbb}), followed by the decay $N\to tbb$. {\bf Right}:
process in the UV complete model, pair production of color-triplet scalars $T, T^*$, and followed by $T\to N \bar b$, $T^*\to N b$, and $N\to t bb$ (via virtual $T$ in parentheses). Baryon number and top quark number are broken when both RHNs decay into top quarks using a Majorana mass insertion.}
\label{fig:sstop}
\end{figure*}

An immediate consequence of Eq.~(\ref{Ntbb}) is the possible 
production of same-sign top quarks 
at the LHC and future hadron colliders, due to 
the Majorana nature of RHNs, as shown in the left panel of Fig.~\ref{fig:sstop}.
In this process, the RHN $N_a$ and a top quark are first produced, 
and then $N_a$ decays into another top quark and two bottom quarks.
Because it is a Majorana particle, an on-shell $N_a$ 
is equally likely to decay into $tbb$ or $\bar t \bar b \bar b$ final states.
The violation of baryon number is manifested in terms of the violation of top quark number (by two units).
The sign of the top quark can be inferred from its leptonic decays.
For a RHN with a few hundred GeV mass and the effective cutoff scale $\Lambda/\sqrt{\lambda_a}$ of a few TeV, we find that 
the cross section for this process can be as large as $\sim 0.3 ~\text{fb}$ in the LHC Run-II at 13\,TeV. 
The main background for this signal is from $t\bar t b\bar b$ final states with the lepton charge from a top quark decay misidentified, 
which is suppressed by the small misidentification rate~\cite{Chatrchyan:2011jz}.
In Table~\ref{t1}, we list the leading order cross sections of our signal for several sample mass values of RHNs.
These points have not been excluded by the existing LHC data.
For example, with $M_a=200$~GeV and $\Lambda/\sqrt{\lambda_a}=1.5\,$TeV, the cross section at 8 TeV is 0.07\,fb, which implies only 1-2 events given the existing integrated luminosity $\sim 27\,{\rm fb}^{-1}$, and it is further suppressed by the top quark leptonic branching ratios.

\begin{table}[h]
\begin{centering}
\begin{tabular}{|c|c|c|c|c|}
\hline
\multicolumn{5}{|c|}{$\sigma(pp\to t N \to ttbb)$} \\
\hline
$M_a$ & 200 GeV & 500 GeV & 800 GeV & 1 TeV\\
\hline 
$\sqrt{s} = 13$~TeV & 0.34 fb & 0.16 fb & $8\times10^{-2}$ fb & $5\times10^{-2}$ fb\\
\hline 
\end{tabular}
\end{centering}
\caption{Same-sign top quark production cross section, at the 13 TeV LHC, via a Majorana RHN and the contact operators 
in Eq.~(\ref{Ntbb}). The cutoff scale is fixed to be $\Lambda/\sqrt{\lambda_a}=1.5\,$TeV. }\label{t1}
\end{table}

Following the same logic as introducing RHNs to make the SM renormalizable, we now discuss a UV completion that 
generates the effective operator Eq.~(\ref{Ntbb}). Given a TeV scale cutoff, it is possible to directly probe the heavy particles in such a model in LHC Run-II and future hadron colliders. 
The model is an extension of the SM that contains a color-triplet
scalar, $T$, with quantum numbers $(\bar 3,1, 1/3)$. The corresponding Lagrangian is 
\beq
\mathcal{L}_{\rm UV} = f_{a}\, T\,\bar N^c_a P_R b +  f'\,T^*\,\bar t^c P_R b   + M_{T}^2 |T|^2 \ .
\label{UV}
\eeq
In fact, this is the simplest model that yields the flavor and color structures of the effective operators in Eq.~(\ref{Ntbb}), 
after integrating out the color-triplet scalar $T$, corresponding to a cutoff $\lambda_a/\Lambda^2 \equiv f_a f'/M_{T}^2$.
The TeV scale cutoff as discussed above can be naturally obtained for $M_T\sim$~TeV and $f_a f' \sim \mathcal{O}(1)$.  
We note that in the above UV model, it is possible to have baryogenesis through 
the decays of the $T$ particle~\cite{BOM}.  We will not further explore such a possibility in this work.

The introduction of the scalar $T$ could offer richer phenomenology at colliders.
If light enough, $T, T^*$ can be pair produced at hadron colliders. Each triplet will first decay into $N + b$, which is then 
followed by subsequent decay $N\to t\,b\,b$ via a virtual $T$. 
The above chain of processes are represented by the diagram in the right 
panel of Fig.~\ref{fig:sstop}. These together result in same-sign top quark final states with many $b$-jets. 
In Table~\ref{t2}, we give the leading-order QCD cross section for the $T, T^*$ pair production at the 13 TeV LHC and a 
100 TeV proton-proton collider, calculated with MadGraph~\cite{Alwall:2011uj}.

\begin{table}[h]
\begin{centering}
\begin{tabular}{|c|c|c|c|c|}
\hline
\multicolumn{5}{|c|}{$\sigma(pp\to TT^*)$} \\
\hline
$M_T$ & \,\,1.5 TeV\,\, & \,\,2 TeV\,\, & \,\,5 TeV\,\, & 10 TeV\\
\hline 
$\sqrt{s} = 13$~TeV & 0.16 fb & 0.01 fb & --- & ---\\
\hline 
$\sqrt{s} = 100$~TeV & 384 fb & 92 fb & 0.54 fb & $4\times10^{-3}$ fb\\
\hline 
\end{tabular}
\end{centering}
\caption{Pair production cross sections of $T, T^*$ via strong interaction at the 13 and 100 TeV proton-proton colliders.}\label{t2}
\end{table}

Moreover, an additional distinct signal could be displaced 
vertices from the decay of RHNs, if we take a somewhat larger cutoff scale 
$\Lambda/\sqrt{\lambda_a}$.   In fact, we find for $M_1 =200$~GeV 
and $\Lambda/\sqrt{\lambda_1}\gtrsim7$~TeV, \eq{GamN1} implies a displaced decay length 
$c\tau_{N_1} \gtrsim 100$~$\mu$m, which would be detectable at the LHC~\cite{Han:2005mu}.
This could result from \eq{UV} for $M_T \sim 1-2$~TeV and $f_a \sim f' \sim 0.2$.  
Events with same-sign tops and displaced vertices would be quite striking 
and hard to miss in collider experiments.
Meanwhile, if the corresponding neutrino Yukawa coupling of $N_1$ is $y_N^1\gtrsim 10^{-7}$, sufficient to explain the solar neutrino mass difference~\cite{Agashe:2014kda}, the partial decay rate of $N_1\to W\ell$ can be as large as order one.
The leptonic decays can be used to identify $N_1$ as a RH neutrino (see, {\it e.g.}, \cite{Nemevsek:2012iq}).

We note that the $T$ particle introduced above has the same quantum numbers as the color-triplet partner of a Higgs doublet in the fundamental ${\bf 5}_H$ representation of the SU(5) 
GUT~\cite{Georgi:1974sy}. This gives the motivation to consider a more unified framework for our scenario~\cite{susy}.
Our light color-triplet $T$ cannot arise from the same ${\bf 5}_H$ as the SM Higgs, 
whose Yukawa couplings are inconsistent with those in \eq{UV}, within our framework. 
One could introduce a new $\bar {\bf 5}_H' = (T, D)$ scalar, where $D$ is a Higgs doublet with quantum numbers $(1,2,-1/2)$ under the SM.
The $T$ couplings in \eq{UV} can then come from the SU(5) gauge invariant Yukawa interactions,
\begin{eqnarray}
\mathcal{L}_{\rm SU(5)} &=& f_{a} {\bf 5}_3  {\bf 1}_a \bar{\bf 5}'_H + M_T^2 \bar{\bf 5}'_H {\bf 5}'_H \nonumber \\ \
&+& f_0\, \overline{\bf 10}_3 {\bf 5}_3 {\bf 5}'_H ({{\bf 24}_H+v_{\rm GUT}})/{\Lambda_{\rm GUT}}\ ,
\label{GUT}
\end{eqnarray}
where ${\bf 1}_a$ are RHNs and singlets under SU(5), the lower index 3 means only the third generation fermions are involved, and the bar over a representation means the complex conjugate of it. 
The last term contains a higher dimensional operator which after GUT symmetry breaking, $\langle{\bf 24}_H \rangle =v_{\rm GUT}\, {\rm diag}(2/3, 2/3, 2/3, -1, -1)$, projects out the $\bar t^c P_R b T^*$ operator in \eq{UV} 
with $f'=5f_0 v_{GUT}/(3\Lambda_{\rm GUT})$.  
At the same time it forbids the dangerous operator $\bar Q L T$, thus evading the usual doublet-triplet splitting problem~\cite{Dvali:1995hp}. 
As a consequence of this setup, the quark Yukawa interaction $\bar Q b_R D$ of the SU(2) Higgs doublet $D$ 
is also forbidden; it still possesses the neutrino Yukawa interaction from the $f_{a}$ term.  Without further splitting of the $T$ and $D$ components, 
the mass of $D$ also lies at the TeV scale with a leptophilic nature.

\bigskip
\noindent{\it Acknowledgments.}
We would like to thank B. Dev, P. Meade, R. Mohapatra, and G. Senjanovi\'{c} for discussions.  We also 
thank B. Dev and R. Mohapatra for informing us of their forthcoming paper on related topics \cite{DM}.   
The work of H.D. is supported in part by the United States Department of Energy
under Grant Contracts DE-SC0012704. The work of Y.Z. is supported by the Gordon and Betty Moore Foundation through Grant \#776 to the Caltech Moore Center for Theoretical Cosmology and Physics, and by the DOE Grant DE-FG02-92ER40701, and also by a DOE Early Career Award under Grant No. DE-SC0010255.
Y.Z. thanks the BNL theory group for hospitality at the final stage of this work. 




\begin{thebibliography}{99}
\bibitem{exception}
The only exception is the operator involving the charm quark $[\bar N_a^c P_R b][\bar c^c P_R b]$ which we find can also have a $\sim$\,TeV scale cutoff.
We focus on the operator \eq{Ntbb} which gives richer collider phenomenology.
\bibitem{susy}
{Another avenue might be to identify $T$ as the supersymmetric partner of a down-type quark. However, holomorphicity dictates $T$ cannot be the superpartner of the existing 
down-type quarks, $d, s, b$, but a new vector-like down-type quark, when promoting the $f_a$ term in \eq{UV} to an R-parity violating superpotential. Thus one has to go further beyond the MSSM, which is less appealing to the authors.}
\end{thebibliography}
\end{document}